\documentclass{iau} 
\usepackage{graphicx}
\usepackage{subcaption}
\usepackage{aas_macros}

\title[The dynamo-wind feedback loop] 
{The dynamo-wind feedback loop : Assessing their non-linear interplay}

\author[Barbara Perri, Allan Sacha Brun, Antoine Strugarek \& Victor R{\'e}ville]   
{Barbara Perri$^{1,2}$, Allan Sacha Brun$^1$, Antoine Strugarek$^1$, 
 \and Victor R{\'e}ville$^3$}

\affiliation{$^1$AIM, CEA, CNRS,
Universit{\'e} Paris-Saclay, Universit{\'e} Paris-Diderot, Sorbonne Paris Cit{\'e},
F-91191 Gif-sur-Yvette, France \\[\affilskip]
$^2$Institut d'Astrophysique Spatiale, CNRS, Universit{\'e} Paris-Sud, Universit{\'e} Paris-Saclay, Bât. 121, 91405 Orsay Cedex, France \\[\affilskip]
$^3$IRAP, Universit{\'e} de Toulouse, CNRS, UPS, CNES, Toulouse, France, 14 Avenue Edouard Belin, F-31400 Toulouse, France}

\pubyear{2019}
\volume{354}  
\jname{Solar and stellar magnetic fields : \\ Origins and manifestations}
\editors{A. Kosovichev, K. Strassmeier \& M. Jardine, eds.}

\begin{document}
\raggedbottom

\maketitle

\begin{abstract}
Though generated deep inside the convection zone, the solar magnetic field has a direct impact on the Earth space environment via the Parker spiral. It strongly modulates the solar wind in the whole heliosphere, especially its latitudinal and longitudinal speed distribution over the years. However the wind also influences the topology of the coronal magnetic field by opening the magnetic field lines in the coronal holes, which can affect the inner magnetic field of the star by altering the dynamo boundary conditions. This coupling is especially difficult to model because it covers a large variety of spatio-temporal scales. Quasi-static studies have begun to help us unveil how the dynamo-generated magnetic field shapes the wind, but the full interplay between the solar dynamo and the solar wind still eludes our understanding.

We use the compressible magnetohydrodynamical (MHD) code PLUTO to compute simultaneously in 2.5D the generation and evolution of magnetic field inside the star via an $\alpha$-$\Omega$ dynamo process and the corresponding evolution of a polytropic coronal wind over several activity cycles for a young Sun. A multi-layered boundary condition at the surface of the star connects the inner and outer stellar layers, allowing both to adapt dynamically. Our continuously coupled dynamo-wind model allows us to characterize how the solar wind conditions change as a function of the cycle phase, and also to quantify the evolution of integrated quantities such as the Alfv{\'e}n radius. We further assess the impact of the solar wind on the dynamo itself by comparing our results with and without wind feedback.

\keywords{(magnetohydrodynamics:) MHD, Sun: activity, Sun: corona, Sun: magnetic fields, (Sun:) solar wind, (Sun:) solar-terrestrial relations}
\end{abstract}

\firstsection 
\section{Introduction}

The Sun exhibits a magnetic activity cycle, which has an 11-year period for amplitude. One general framework explaining such a generation of large-scale magnetic field is the interface dynamo (\cite[Parker 1993]{Parker1993}): the differential rotation profile in the convection zone of the star (\cite[Thompson et al. 2003]{Thompson2003}) leads to the generation of strong toroidal fields at the tachocline (\cite[Spiegel \& Zahn 1992]{Spiegel1992}), which in turn is used to regenerate poloidal fields thanks to the combination of turbulence, buoyancy and Coriolis force at the surface. This dynamo loop allows for the amplification of the initial magnetic field until saturation, thus sustaining it against ohmic dissipation (\cite[Brun \& Browning 2017]{BrunBrowning2017}). A simplified yet efficient approach to model it is the mean-field dynamo framework (\cite[Moffatt 1978]{Moffatt1978}), focusing on large-scale fields and assuming axisymmetry. The generation of toroidal field through differential rotation is then deemed the $\Omega$ effect, and the regeneration of the poloidal or toroidal field via turbulence is deemed the $\alpha$ effect. This description has the advantages of being easy to implement in MHD simulations with low computational costs, and yielding realistic results (\cite[Charbonneau 2010]{Charbonneau2010}). On the other hand, in the outer layers of the Sun, one of the main phenomena is the transsonic and transalfv{\'e}nic solar wind. The first hydrodynamical description was given by \cite[Parker (1958)]{Parker1958}, and magnetism was added by \cite[Weber \& Davis (1967)]{Weber1967} and \cite[Sakurai (1985)]{Sakurai1985} to yield a better description of the corresponding torque applied to the star. Stellar wind of solar-like stars can be described using 2.5D axisymmetric MHD simulations as well (\cite[Keppens \& Goedbloed 1999]{Keppens1999}; \cite[Matt \& Pudritz 2008]{Matt2008}).

Recent observations, by the satellite \textit{Ulysses} for cycle 22 and the beginning of cycle 23 (\cite[McComas et al. 2008]{McComas2008}), or by the satellite \textit{OMNI} for cycles 23 and 24 (\cite[Owens et al. 2017]{Owens2017}), have shown that there is a correlation between the 11-year dynamo cycle and the evolution of the corona. During a minimum of activity, the magnetic field is low in amplitude and its topology is mostly dipolar ; the corona is very structured with fast wind at the poles (around 800 km/s) associated with coronal holes, and slow wind at the equator (around 400 km/s) associated with streamers. During a maximum of activity, the magnetic field is high in amplitude and its topology is a mixture of high modes, dominated by the quadrupolar modes (\cite[DeRosa et al. 2012]{DeRosa2012}) ; fast and slow solar winds can be found at all latitudes. This suggests that there is a coupling operating between the interior and the exterior of the Sun, but we still don't know precisely how it is operating and on which timescales. From a theoretical and numerical point of view, it is however very difficult to study all of these layers simultaneously : the magnetic field and the wind evolve over very different scales (from hours to years and from a few solar radii to 1 AU). The physical properties of their respective environment are also very different ; take for instance the rapidly changing $\beta$ plasma parameter, which is the ratio of the thermal pressure over the magnetic pressure, from more than 1 inside the star to less than 1 in the chromosphere (\cite[Gary 2001]{Gary2001}). Finally, from a mathematical point of view, the MHD equations are stiff, meaning that it requires small time and grid steps to be solved. All of these disparities make the modeling of this coupling a numerical challenge.

There have been various attempts to resolve this problem with different approaches. A first attempt is to use a quasi-static approach, meaning that the coupling is modeled through a series of wind relaxed states corresponding to a sequence of magnetic field configurations evolving in time. These models can be data-driven, using series of magnetic field observations (\cite[Luhmann et al. 2002]{Luhmann2002}; \cite[R{\'e}ville \& Brun 2017]{Reville2017}), or rely on the numerical coupling between two codes dedicated respectively to the inner and outer layers of the Sun (\cite[Pinto et al. 2011]{Pinto2011}; \cite[Perri et al. 2018]{Perri2018b}). Another approach is to zoom on the surface with a numerical box of a few tens of Megameters to capture the small time and spatial scales, which means this approach can include small-scale physical processes (for example convection or radiative transfer at the surface) but on a short period of time and only for a specific region of the Sun (\cite[V{\"o}gler et al. 2005]{Vogler2005}; \cite[Stein \& Nordlund 2006]{Stein200}; \cite[Wedemeyer-B{\"o}hm et al. 2009]{Wedemeyer2009}; \cite[Gudiksen et al. 2011]{Gudiksen2011}). Finally there have been some attempts to model a dynamical coupling on a global scale, for example in \cite[von Rekowski \& Brandenburg (2006)]{VonRekowski2006} with a simulation box including both the star and its corona, but for a T-Tauri star and with a disk interaction. Our aim is to focus on solar-like stars and on the large-scale field by using mean-field and axisymmetry assumptions. We design a 2.5D numerical model including the star and its corona, and design an interface to control the complex and diverse interactions between the two zones.  

\firstsection 
\section{Numerical setup}

\subsection{Wind model}

Our wind model is adapted from \cite[R{\'e}ville et al. (2015)]{Reville2015a}. We solve the set of the conservative ideal MHD equations composed of continuity equation for the density $\rho$, the momentum equation for the velocity field $\mathbf{v}$ with its momentum written $\mathbf{m}=\rho\mathbf{v}$, the equation for the total energy $E$ and the induction equation for the magnetic field $\mathbf{B}$:
\begin{equation}
\frac{\partial}{\partial t}\rho+\nabla\cdot\rho\mathbf{v}=0,
\end{equation} 
\begin{equation}
\frac{\partial}{\partial t}\mathbf{m}+\nabla\cdot(\mathbf{mv}-\mathbf{BB}+\mathbf{I}p) = \rho\mathbf{a},
\end{equation}
\begin{equation}
\frac{\partial}{\partial t}E + \nabla\cdot((E+p)\mathbf{v}-\mathbf{B}(\mathbf{v}\cdot\mathbf{B})) = \mathbf{m}\cdot\mathbf{a},
\end{equation}
\begin{equation}
\frac{\partial}{\partial t}\mathbf{B}+\nabla\cdot(\mathbf{vB}-\mathbf{Bv})=0,
\label{eq:induction_pluto_wind}
\end{equation}
where $p$ is the total pressure (thermal and magnetic), $\mathbf{I}$ is the identity matrix and $\mathbf{a}$ is a source term (gravitational acceleration in our case). We use the ideal equation of state $\rho\varepsilon = p_{th}/(\gamma -1)$, where $p_{th}$ is the thermal pressure, $\varepsilon$ is the internal energy per mass and $\gamma$ is the adiabatic exponent. This gives for the energy : $E = \rho\varepsilon+\mathbf{m}^2/(2\rho)+\mathbf{B}^2/2$.

PLUTO solves normalized equations, using three variables to set all the others: length, density and speed. If we note with $*$ the parameters related to the star and with $0$ the parameters related to the normalization, we have $R_*/R_0=1$, $\rho_*/\rho_0=1$ and $V_{K}/V_0=\sqrt{GM_*/R_*}/V_0=1$, where $V_{K}$ is the Keplerian speed at the stellar surface and $G$ the gravitational constant. In our set-up, we choose $R_0=R_\odot=6.96 \ 10^{10}$ cm, $\rho_0=\rho_\odot=6.68 \ 10^{-16} \ \mathrm{g/cm}^3$ and $V_0=V_{K,\odot}=4.37 \ 10^2$ km/s. Our wind simulations are then controlled by three parameters : the adiabatic exponent $\gamma=1.05$ for the polytropic wind, the rotation of the star normalized by the escape velocity $v_{rot}/v_{esc}=2.93 \ 10^{-3}$ and the speed of sound normalized also by the escape velocity $c_s/v_{esc}=0.243$. Note that the escape velocity is defined as $v_{esc} = \sqrt{2}V_{K} = \sqrt{2GM_*/R_*}$. Such values correspond to a $1.3 \ 10^6$ K hot isothermal corona rotating at the solar rotation rate. 

We assume axisymmetry and use the spherical coordinates $(r,\theta,\phi)$. We choose a finite-volume method using an approximate Riemann Solver (here the HLL solver, cf. \cite[Einfeldt (1988)]{Einfeldt1988}). PLUTO uses a reconstruct-solve-average approach using a set of primitive variables $(\rho,\mathbf{v},p,\mathbf{B})$ to solve the Riemann problem corresponding to the previous set of equations. The time evolution is then implemented via a second order Runge-Kutta method. To enforce the divergence-free property of the field, we use a hyperbolic divergence cleaning, which means that the induction equation is coupled to a generalized Lagrange multiplier in order to compensate the deviations from a divergence-free field (\cite[Dedner 2002]{Dedner2002}). We do not use the traditional approach of splitting between the curl-free background field and the fluctuation field $\delta \mathbf{B}$, because in our case the background field will be the dynamo field generated inside the star and evolving at each time-step.

\begin{figure}
    \centering
    \includegraphics[width=0.6\textwidth]{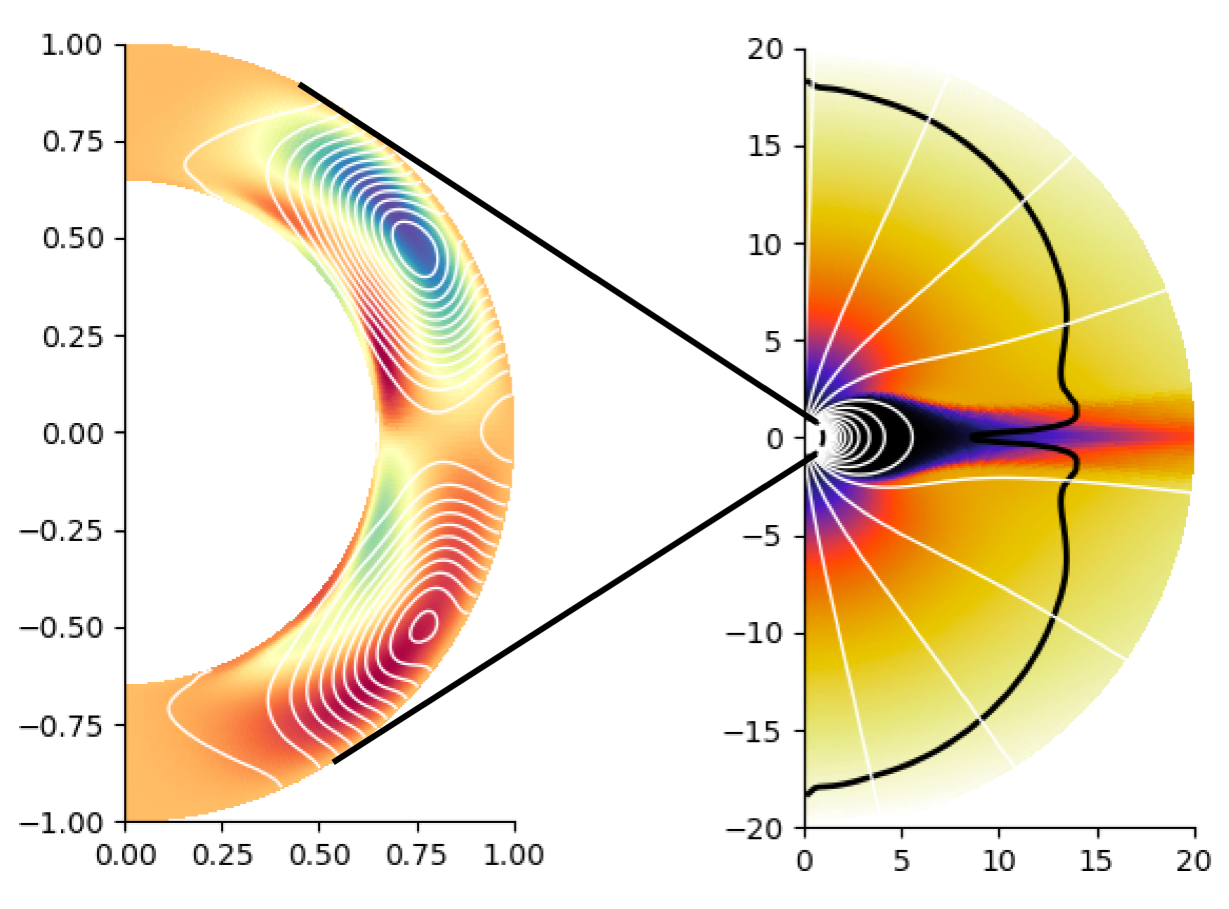}
    \caption{Examples of simulations with the dynamo (on the left) or the wind (on the right) models. For the dynamo the colorscale represents the toroidal magnetic field $B_\phi$ and the white lines are the poloidal magnetic field lines. For the wind the colorscale represents the quantity $\mathbf{v}\cdot\mathbf{B}/(c_s||\mathbf{B}||)$, the black line corresponds to the Alfv{\'e}n radius and the white lines to the poloidal magnetic field lines.}
    \label{fig:dyn_wind}
\end{figure}

The numerical domain dedicated to the wind computation is an annular meridional cut with the colatitude $\theta \in [0,\pi]$ and the radius $r \in [1.01,20]R_\odot$. We use an uniform grid in latitude with 256 points, and a stretched grid in radius with 400 points; the grid spacing is geometrically increasing from $\Delta r/R_*=0.002$ at the surface of the star to $\Delta r/R_*=0.02$ at the outer boundary. At the latitudinal boundaries ($\theta=0$ and $\theta=\pi$), we set axisymmetric boundary conditions. At the top radial boundary ($r=20 R_*$), we set an outflow boundary condition which corresponds to $\partial/\partial r=0$ for all variables, except for the radial magnetic field where we enforce $\partial(r^2B_r)/\partial r=0$. Because the wind has opened the field lines and under the assumption of axisymmetry, this ensures the divergence-free property of the field. We initialize the velocity field with a polytropic wind solution and the magnetic field with a dipole. The right panel of Figure \ref{fig:dyn_wind} shows an example of a wind simulation only.

\subsection{Dynamo model}

As we have seen in equation \ref{eq:induction_pluto_wind}, PLUTO solves the full non-linear ideal induction equation. For the dynamo inside the star, we will consider that the magnetic field $\mathbf{B}$ is the large-scale mean field, and thus implement an alternative form of the induction equation with ohmic diffusion:
\begin{equation}
\frac{\partial}{\partial t}\mathbf{B}+\nabla\cdot(\mathbf{vB}-\mathbf{Bv})=\nabla\times\left(\alpha\mathbf{B}\right) - \nabla\times\left(\eta\times\nabla\times\mathbf{B}\right),
\end{equation}
where $\eta$ is the effective magnetic diffusivity and $\alpha$ is a coefficient for the $\alpha$ effect obtained by First Order Smooth Approximation (FOSA) of the electro-motive force (emf) (\cite[Pouquet et al. 1976]{Pouquet1976}). Hence the $\Omega$ effect is taken into account with the second term on the left, and the $\alpha$ effect with the first one on the right. This form of the induction equation is only active inside the star ($r<R_\odot$); no other equation is solved there. This new induction equation follows the same normalization as described before. However, when talking about dynamo parameters, the community usually refers to the parameters $C_\alpha = \alpha_0R_\odot/\eta_t$, $C_\Omega=\Omega_0R_\odot^2/\eta_t$ and $R_e=V_0R_\odot/\eta_t$. To make it more convenient, we will use in this article the traditional control parameters of the dynamo models, just note that there is a factor $\eta_t/(R_\odot V_K)$ to switch to the PLUTO normalization (where $\eta_t$ is the turbulent magnetic diffusivity, see eq. \ref{eq:eta}).

For the physical parameters, we got inspiration from case B of \cite[Jouve et al. (2008)]{Jouve2008}. The rotation in this zone is solar-like with a solid body rotation below $0.66R_\odot$ and differential rotation above, with the equator rotating faster than the poles:
\begin{equation}
\Omega(r,\theta) = \Omega_c + \frac{1}{2}\left(1+\rm{erf}\left(\frac{r-r_c}{d}\right)\right)\left(1-\Omega_c-c_2\rm{cos}^2\theta\right),
\end{equation}
where $\Omega_c = 0.92$, $r_c=0.7R_\odot$, $d = 0.02$ and $c_2 = 0.2$. We recall that the physical amplitude of the $\Omega$ effect is given by the $C_\Omega$ parameter. In this model, we do not have any poloidal flows, and hence no meridional circulation.

As a first simple approximation, the $\alpha$ effect has no latitudinal and radial dependence in the convection zone, and is zero in the radiative zone, with a smooth transition between the two zones :
\begin{equation}
\alpha(r,\theta) = \frac{3\sqrt{3}}{4}\,\rm{sin}^2\theta\,\rm{cos}\,\theta\left(1+\rm{erf}\left(\frac{r-r_c}{d}\right)\right) \left(1+\left(\frac{B_\phi(r_c,\theta,t)}{B_0}\right)^2\right)^{-1}.
\label{eq:alpha_effect}
\end{equation}
The factor $3\sqrt{3}/4$ is used as a normalization to have a maximum amplitude of 1 in the convection zone for this profile. The quenching term (which is the last term in equation \ref{eq:alpha_effect}) allows for saturation of the magnetic field near the reference magnetic field value $B_0$.

We have a jump in diffusivity between the radiative and the convection zone of two orders of magnitudes:
\begin{equation}
\frac{\eta}{\eta_t}(r) = \frac{\eta_c}{\eta_t} + \frac{1}{2}\left(1-\frac{\eta_c}{\eta_t}\right)\left(1+\rm{erf}\left(\frac{r-r_c}{d}\right)\right),
\label{eq:eta}
\end{equation}
with $\eta_c/\eta_t=10^{-2}$ and $\eta_t$ a parameter that we fix. The magnetic field is initialized with a dipole confined in the convection zone.
Hence $B_\phi$ is initially equal to 0, but will grow through dynamo action.

The numerical domain dedicated to the dynamo computation is an annular meridional cut with the colatitude $\theta \in [0,\pi]$ and the radius $r \in [0.6,1.01]R_\odot$. We use a uniform grid in latitude with 256 points, and a uniform grid in radius with 200 points, which yields a grid spacing of $\Delta r/R_*=0.002$. At the latitudinal boundaries ($\theta=0$ and $\theta=\pi$), we set axisymmetric boundary conditions. For the bottom boundary condition ($r=0.65R_\odot$) we use a perfect conductor condition.
For the top boundary condition ($r=R_\odot$), we use the two first layers of the interface, which will be described in the next section. The left panel of Figure \ref{fig:dyn_wind} shows an example of a dynamo simulation only.

\subsection{Interface principles}

\begin{figure}
    \centering
    \includegraphics[width=\textwidth]{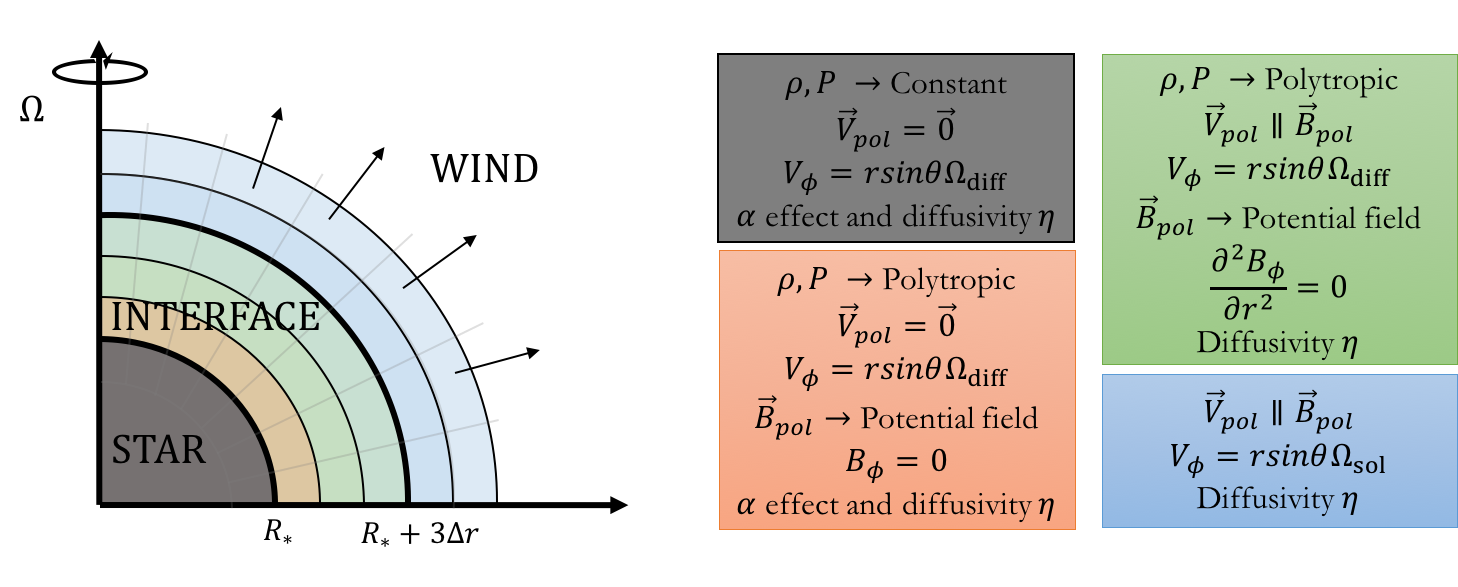}
    \caption{Schematic of our interface region. The grey box describes the conditions for inside the star, the orange one for the first layer of the interface, the green one for the last two points of the interface and the blue one for the first two points of the wind model.}
    \label{fig:bc_coupling}
\end{figure}

The interface layer between the dynamo and the wind computational zones is shown in Figure \ref{fig:bc_coupling} by the orange and green areas. This interface is crucial to control and understand the interactions between the two zones.

The interface is divided into three layers of one radial grid point for all latitudes. This is tailored to a numerical method using a 2-point stencil linear reconstruction, but could be adapted to other methods. The first two layers constitute the boundary condition for the dynamo, and the last two layers constitute the boundary condition for the wind. We also alter the solution in the first two points of the wind computational domain for more stability. The first layer in orange is very similar to the dynamo zone, except that now the density and pressure decrease following a polytropic law (but with a continuous link with the constant value inside the star), and the magnetic field is extrapolated as a potential field using the value of the last point of the dynamo zone. We still have an $\alpha$ effect for continuity, but the equations are not evolved in the interface. In the second and third layers of the interface in green, we set conditions for the wind : the poloidal speed is aligned with the poloidal magnetic field extrapolated from below and $\partial_r^2B_\phi$ is set to 0 to limit the generation of currents at the surface of the star (imposed from right to left). In the first two layers of the wind computational domain in blue, we impose the poloidal speed to remain parallel to the wind poloidal field to limit again the generation of currents. We also impose some diffusivity a bit further than the star surface using the following expression :
\begin{equation}
\eta = \frac{1}{2\eta_{norm}}\left(\eta_c+\frac{1}{2}(\eta_t-\eta_c)\right)\left(1+\rm{erf}\left(\frac{r-r_c}{d}\right)\right)\left(1+\rm{tanh}\left(\frac{r_\eta-r}{d_\eta}\right)\right),
\label{eq:eta_profile}
\end{equation}
with $r_\eta = 1.015$ and $d_\eta = 0.003$. This allows the diffusivity to drop only after around ten grid points above the interface. The way the interface is designed, the dynamo magnetic field can influence the wind via the potential extrapolation of $B_r$ and $B_\theta$, and the wind can back-react on the dynamo via the $B_\phi$ condition that changes the dynamo boundary conditions. 

\firstsection
\section{Evolution along activity cycles}

\begin{figure}
    \centering
    \begin{subfigure}[t]{0.45\textwidth}
    \includegraphics[width=\textwidth]{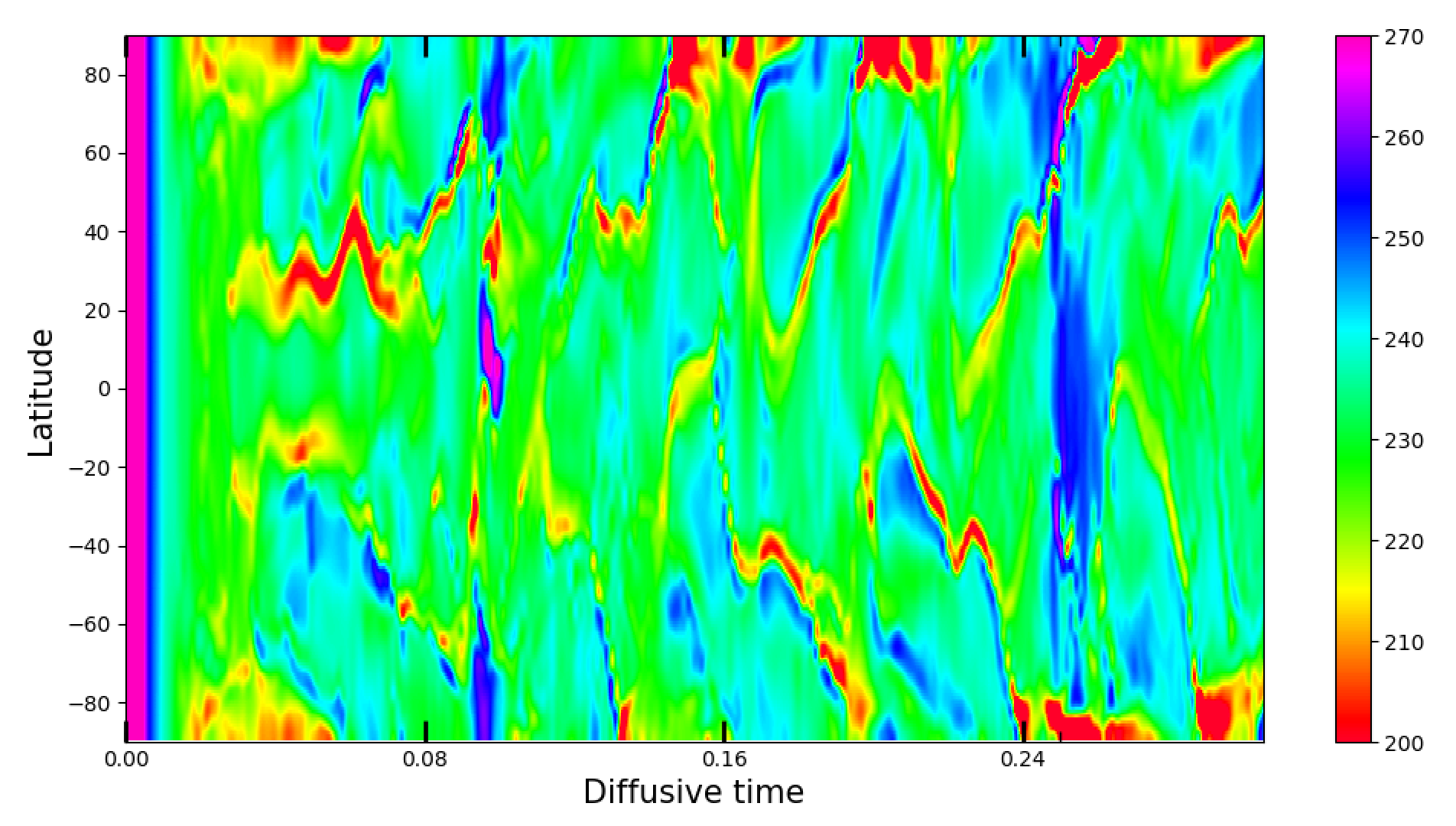}
    \end{subfigure}
    \begin{subfigure}[t]{0.42\textwidth}
    \includegraphics[width=\textwidth]{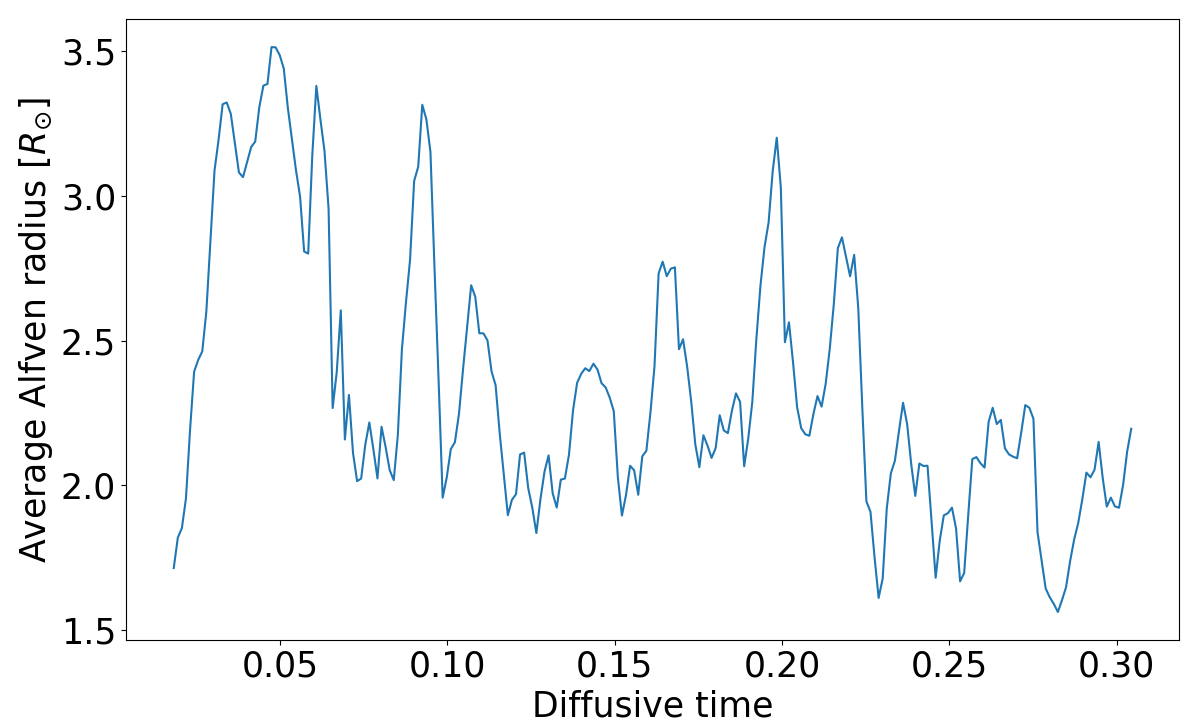}
    \end{subfigure}
    \caption{Evolution of the wind in the coupled model. On the left we show a time-latitude diagram of the wind speed at $20R_\odot$ in km/s. On the right we show the time evolution of the average Alfv{\'e}n radius as a fraction of the diffusive time $t_\eta=R_\odot^2/\eta_t$.}
    \label{fig:wind_evol_coupling}
\end{figure}

To validate the coupling from a theoretical point of view, we focus on a model whose dynamo period is shorter than the solar dynamo, thus more related to a young Sun. This helps bring closer the timescales of the dynamo and the wind. To design this model, we adapt the magnetic diffusivity to set the cycle period by adjusting the diffusive time $t_\eta = R_\odot^2/\eta_t$ using $\eta_t=3.7 \ 10^{14} \ \rm{cm}^2.\rm{s}^{-1}$; then we set $C_\Omega=3.4 \ 10^1$ to have the solar rotation rate ; finally we set $C_\alpha=1.44 \ 10^4$ to have a dynamo number $D=C_\Omega \times C_\alpha$ above the dynamo threshold. This yields a case where $C_\Omega$ is smaller than $C_\alpha$, which corresponds to a strong generation of poloidal field due to convective turbulence.

This model allows us to obtain in one simulation and in a completely self-consistent way without breaking causality the dynamical evolution of the corona in response to activity cycles. In Figure \ref{fig:wind_evol_coupling}, we can see the time-latitude diagram for the wind speed. Extrapolated to 1 AU, we obtain speeds between 420 and 560 km/s, which correspond to the slow wind component due to our polytropic approximation. However we see faster and slower components inside our wind with a difference of up to 70 km/s between the two components. The evolution of the corona is highly dynamical with a lot of transients associated to the continuous response of the wind to the changing field. The associated streamers are very thin and evolve quickly, the wind has the time to adapt to the oscillatory dynamo field but cannot reach a stationary state.

We can also see the evolution of integrated quantities, as shown in the right panel of Figure \ref{fig:wind_evol_coupling}. We will focus on the average Alfv{\'e}n radius $\langle r_A\rangle$. The Alfv{\'e}n radius is defined as the distance at which the wind speed equals the Alfv{\'e}n speed $v_A=B/\sqrt{4\pi\rho}$, and the average Alfv{\'e}n radius, as defined in \cite[Pinto et al. (2011)]{Pinto2011}, corresponds to the Alfv{\'e}n radius averaged by the mass flux through the surface of a sphere. The average Alfv{\'e}n radius evolves between 1.5 and 3.5 $R_\odot$ ; this is smaller than for the present Sun because the coupling needs a weaker field to operate with such a fast dynamo. The mass loss evolves between 3.5 and 6.5 $10^{-14} \ M_\odot/\rm{yr}$, which is just a bit more than the solar values estimated between 2.3 and 3.1 $10^{-14} \ M_\odot/\rm{yr}$ (\cite[McComas et al. 2008]{McComas2008}; \cite[R{\'e}ville \& Brun 2017]{Reville2017}). The angular momentum loss evolves between 1.0 and 8.0 $10^{29}$ cgs, which is the same order of magnitude as for the Sun. We see a lot of modulations in time with the evolution of the activity cycles, obtained for the first time in a completely auto-coherent way.

\firstsection
\section{Feedback loop : influence of the wind on the dynamo solution}

\begin{figure}
    \centering
    \begin{subfigure}[t]{0.38\textwidth}
    \includegraphics[width=\textwidth]{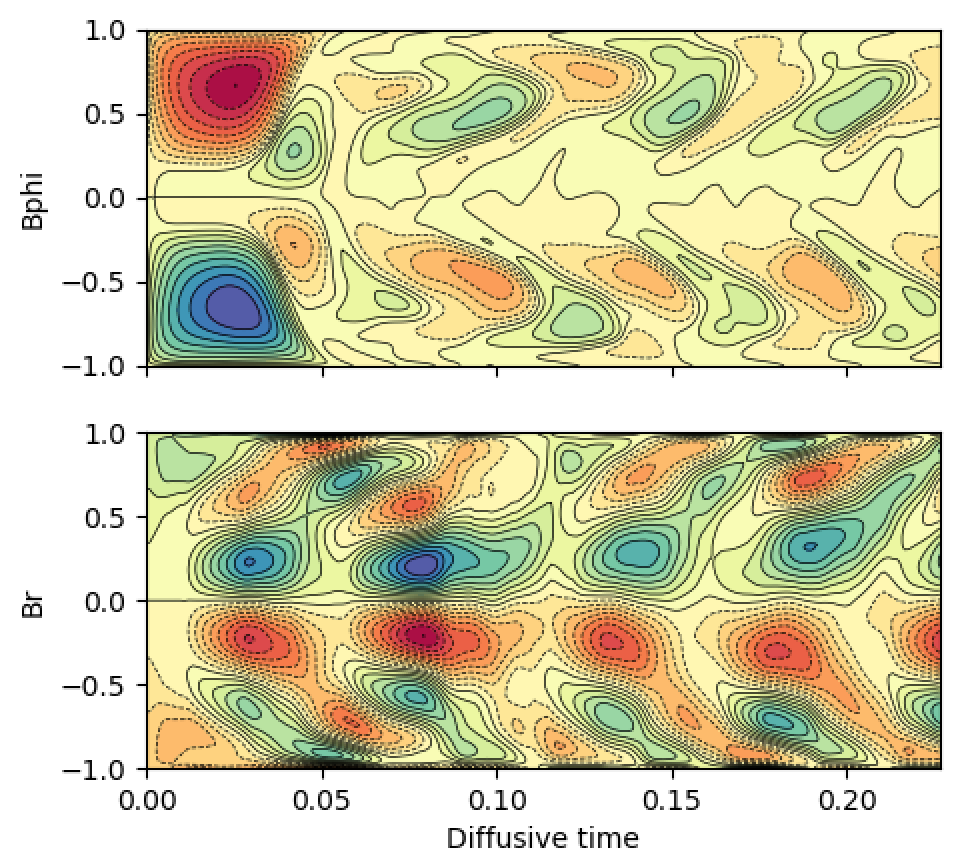}
    \end{subfigure}
    \begin{subfigure}[t]{0.38\textwidth}
    \includegraphics[width=\textwidth]{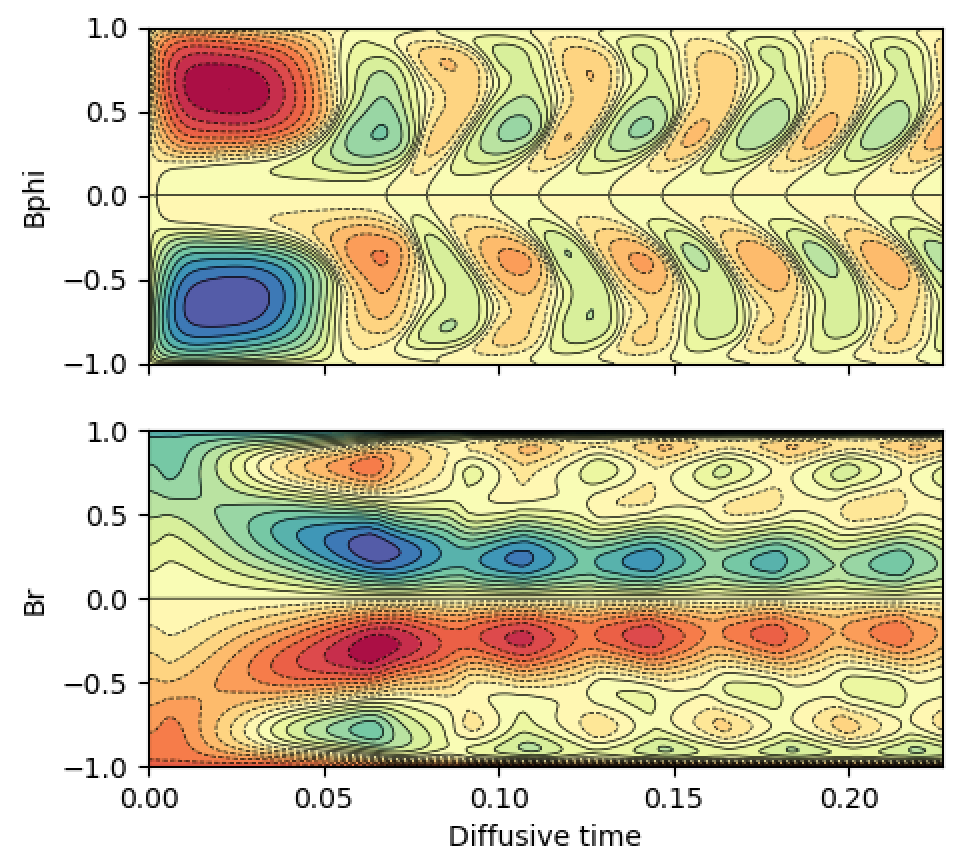}
    \end{subfigure}
    \caption{Butterfly diagrams with and without the feedback of the wind (respectively on the left and on the right).}
    \label{fig:diagpap_bphi}
\end{figure}

\begin{figure}
    \centering
    \begin{subfigure}[t]{0.45\textwidth}
    \includegraphics[width=\textwidth]{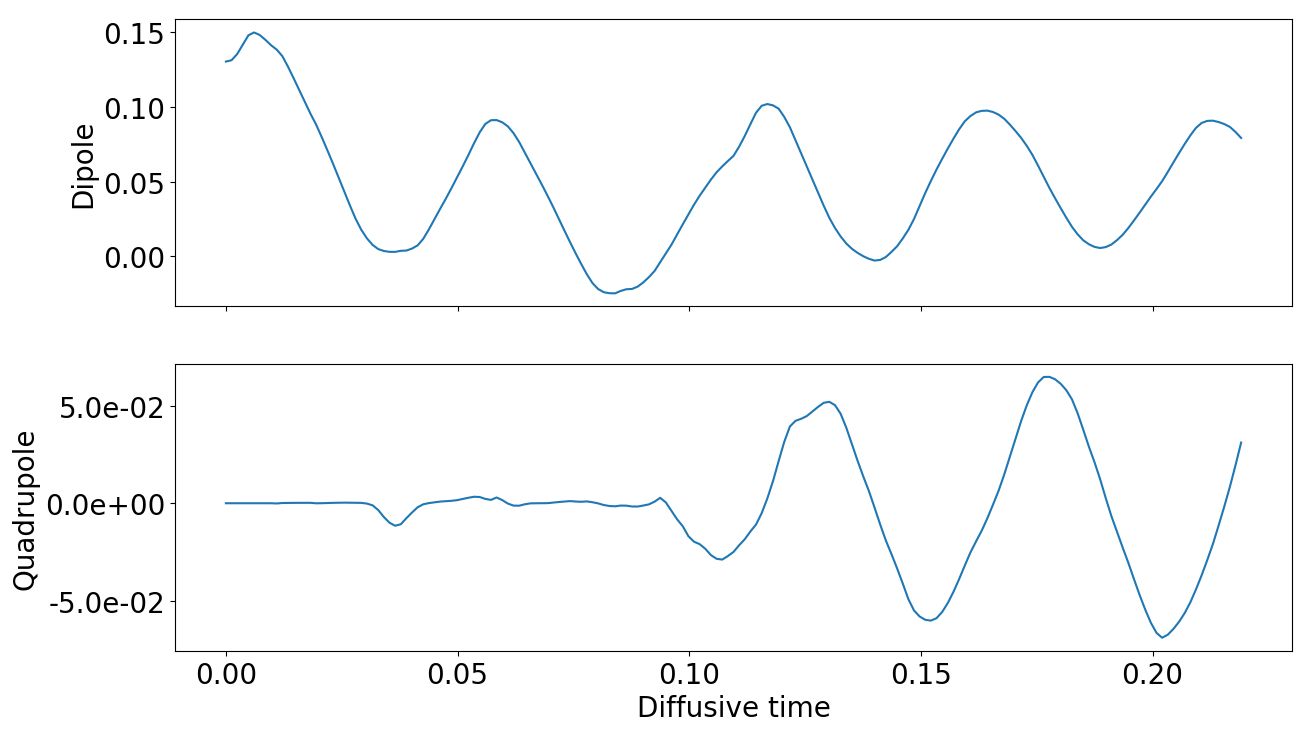}
    \end{subfigure}
    \begin{subfigure}[t]{0.45\textwidth}
    \includegraphics[width=\textwidth]{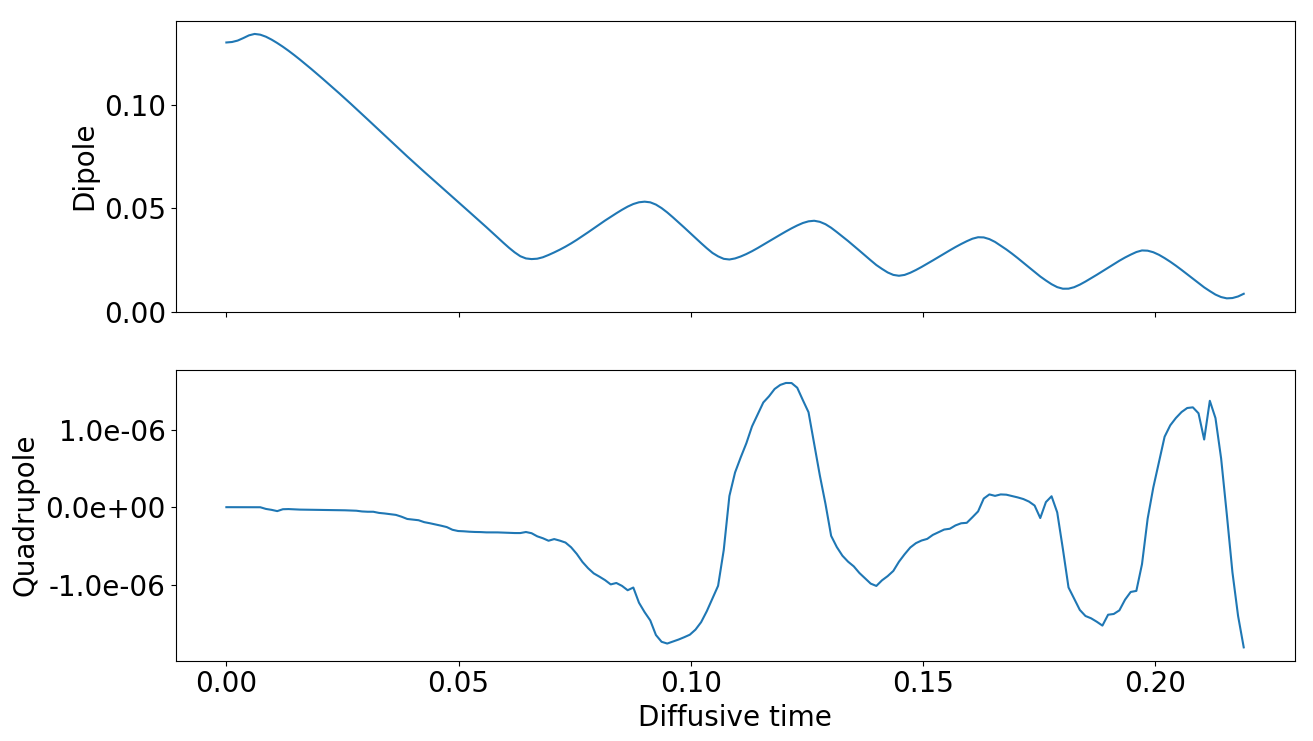}
    \end{subfigure}
    \caption{Time evolution of the dipolar ($\ell=1$) and quadrupolar ($\ell=2$) modes of the surface radial magnetic field with and without the feedback of the wind (respectively on the left and on the right).}
    \label{fig:field_evol_bphi}
\end{figure}

In this last section we will focus on the feedback loop between the dynamo and the wind. As said in the introduction, a variety of codes have shown the influence of the magnetic field generated by the dynamo on the coronal structures, but it is more difficult to evaluate the potential impact of the wind on the interior of the star. To test the impact of this feedback loop, we have run two models : in the first case, we used the interface described before, which allows the wind to back-react on the boundary conditions ; in the second case, we imposed the condition $B_\phi=0$ in the second interface layer, thus cutting the feedback from the wind. The corresponding butterfly diagrams are shown in Figure \ref{fig:diagpap_bphi}. We see a clear difference, starting from $0.03t_\eta$ with $t_\eta=R_\odot^2/\eta_t$. Without the wind influence (right panel), the tachocline toroidal and surface radial magnetic fields are equatorially anti-symmetric and the cycle is regular with a period of $0.05t_\eta$. With the influence of the wind (left panel), the cycle takes a longer time to stabilize to finally reach a period of about $0.05t_\eta$. The cycle tends to become equatorially symmetric, although at most times a North-South asymmetry is still noticeable. To understand this difference, we looked at the evolution in time of the dipolar and quadrupolar modes ($\ell=1$ and $\ell=2$) for these two cases (cf. Figure \ref{fig:field_evol_bphi}). Without the influence of the wind, the dipolar mode is almost constant, slightly decreasing, while the quadrupolar mode has an amplitude 5 orders of magnitude less. Hence the symmetric family is negligible. But with the influence of the wind, the quadrupolar mode grows to an amplitude equivalent of the dipolar mode (70\%). The back-reaction of the wind in this case has thus a visible influence by favoring the growth of the symmetric family by influencing the boundary conditions of the dynamo.

It is too soon to know if this result can be generalized to any coupled system. Indeed, this case has pretty extreme values for a dynamo; \cite[Tavakol et al. (1995)]{Tavakol1995} has shown that in such parameter regimes, the dynamo is highly non-linear and can easily switch from symmetric to anti-symmetric regimes because of the non-linear quenching or asymmetry of the physical parameters. We have also performed a threshold study similar to \cite[Jouve \& Brun (2007)]{Jouve2007} and have determined that for this set of parameters the quadrupolar mode has a growth critical threshold lower than the dipolar mode ($C_\alpha^Q=80$ versus $C_\alpha^D=100$). This study is a highly sophisticated proof of concept to demonstrate that the feedback-loop between the dynamo and the wind is present in simulations, and needs to be more thoroughly investigated to understand the physical implications for stars.

\firstsection

\end{document}